\documentclass{article}
\usepackage[utf8]{inputenc}
\usepackage[affil-it]{authblk}
\usepackage{graphicx}
\usepackage{subfigure}
\usepackage{multirow}
\usepackage{amsmath}
\graphicspath{{figs/}}
\usepackage{xcolor}

\usepackage[margin=2.5cm]{geometry}

\title{Capacitive storage in mycelium substrate}
\author[1]{Alexander E. Beasley\footnote{Corresponding author: Alexander Beasley, alex.beasley@uwe.ac.uk}}
\author[1]{Anna L. Powell}
\author[1]{Andrew Adamatzky}

\affil[1]{Unconventional Computing Laboratory, UWE, Bristol, UK}

\date{\today}

\begin{document}

\maketitle

\begin{abstract}
\noindent 
The emerging field of living technologies aims to create new functional hybrid materials in which living systems interface with artificial ones. Combining research into living technologies with emerging developments in computing architecture has enabled the generation of organic electronics from plants and slime mould. Here, we expand on this work by studying capacitive properties of a substrate colonised by mycelium of grey oyster fungi, \emph{Pleurotus ostreatus}. Capacitors play a fundamental role in traditional analogue and digital electronic systems and have a range of uses including sensing, energy storage and filter circuits. Mycelium has the potential to be used as an organic replacement for traditional capacitor technology. Here, were show that the capacitance of mycelium is in the order of hundreds of pico-Farads. We also demonstrate that the charge density of the mycelium `dielectric' decays rapidly with increasing distance from the source probes. This is important as it indicates that small cells of mycelium could be used as a charge carrier or storage medium, when employed as part of an array with reasonable density.    

\vspace{3mm}

\noindent
\emph{Keywords:} fungi, capacitance, mycelium, storage, biocomputing

    
\end{abstract}

\section{Introduction}
\label{sec:introduction}

The study of novel substrates for sensing, storing and processing information draws on work from the fields of unconventional computing, living technology and organic electronics. The field of unconventional computing aims to define the principles of information processing in living, physical and chemical systems and applies this knowledge to the development of future computing devices and architectures~\cite{advancesUC}. Research into living technologies is focused on the co-functional integration of animate and non-organic systems ~\cite{bedau2010living}. Finally, organic electronics~\cite{shaw2001organic,klauk2006organic} looks to use naturally occurring materials as analogues to traditional semi-conductor circuits~\cite{Stavrinidou2015Plants}, which often requires functionalisation using polymers or metallic compounds to exploit ionic movement~\cite{Leger2008organic}. The development of organic electronics promises a technology that is low-cost and has low production temperature requirements~\cite{book:analogorganicelectronics}. However, difficulties such as relatively low gain of organic transistors (approx. 5), and the behavioural variability inherently means that there is a large amount of research effort being placed into organic thin film transistors~\cite{Zschieschang2019Transistors, Tokito2018TFT, Endoh2007OLIT, Tang2018OTFT}, organic LEDs~\cite{Sano2019OLED, Mizukami2018OLED}, and organic capacitors~\cite{Li2017MOF, Sangermano2015UV, Morimoto1997Capacitor}.
The capacitive properties of a  device allows it to either store energy or react to AC/DC signals differently. There are a number of  applications in which this property may be utilised, such as energy harvesting~\cite{paper:energy_harvesting_capacitors}, memory~\cite{paper:mos_memories}, or filter circuits~\cite{brodersen1979mos}. Hybrid electronic circuits are a concept that looks to combine traditional silicon, semi-conductor devices with elements found in nature~\cite{lu2010nanoelectronics,beausoleil2008nanoelectronic}.

The capacitive properties of living tissues~\cite{mcadams1995tissue} have a wide range of potential applications, e.g. the estimation of a plan root system size~\cite{chloupek1977evaluation,rajkai2002electrical}, quantifying the DNA content of eukaryotic cells~\cite{sohn2000capacitance}, analysing water transport pathways in plants~\cite{blackman2011two}, measuring heat injury in plants~\cite{zhang1993measurement}, measuring contents of minerals in bones~\cite{williams1996electrical}, gauging firmness of apples~\cite{paper:apple_capacitance_firmness}, sugar contents of citrus fruits~\cite{paper:citrus_capacitance_sugar}, maturity of avocados~\cite{book:avacados}, 
estimating depth of epidermal barriers~\cite{boyce1996surface}, studies of endo- and exocytosis of single cells~\cite{rituper2013high}, and approximating mass and morphology of microbial colonies~\cite{fehrenbach1992line, neves2000real,sarra1996relationships}.

In the present study we focus on the capacitive properties of the mycelium of the grey oyster fungi \emph{Pleurotus ostreatus} for several reasons. 

Firstly, research into the capacitive properties of fungi is lacking, despite their huge potential for bioelectronic applications. Fungi are the largest, most widely distributed and oldest group of living organisms on the planet~\cite{carlile2001fungi}. The smallest fungi are microscopic single cells, the largest, \emph{Armillaria bulbosa}, occupies 15 hectares and weighs 10 tons~\cite{smith1992fungus}. Fungi sense light, chemicals, gases, gravity and electric fields~\cite{bahn2007sensing} as well as demonstrating mechanosensing behaviour~\cite{kung2005possible}. Thus, their electrical properties can be tuned via various inputs. 

Secondly, fungi have the potential to be used as distributed living computing devices, i.e. large-scale networks of mycelium, which collect and analyse information about environment and execute some decision making processes~\cite{adamatzky2018towards}. 

Finally, there is a growing interest in developing buildings from pre-fabricated blocks of substrates colonised by  fungi~\cite{ross2016your,appels2019fabrication,dahmen2016soft}. A recent initiative aims to grow monolithic constructions in which living mycelium coexists with dried mycelium, functionalised with nanoparticles and polymers~\cite{adamatzky2019fungal}. In such a case, fungi could be act as optical, tactile and chemical sensors, fuse and process information and perform decision making computations~\cite{adamatzky2018towards}.\par

Providing local charge to areas of mycelium allows the storage of information inside the substrate. Identifying the area around which the induced charge can be detected allows the construction of an array in the substrate where each cell can contain individual bits of information. Determining the capacitive properties of fungi takes a step towards the realisation of fungal analogue circuits --- circuits that use fungi to replace traditional semiconductors. \par

The rest of this paper is organised as follows. Section~\ref{sec:experimental} describes the experimental methods used for the analysis of the substrate. Section~\ref{sec:results} presents the results with discussions. Finally, conclusions are drawn in Sect.~\ref{sec:conclusions}.\par

\section{Experimental method}
\label{sec:experimental}

\begin{figure}[!tpb]
    \centering
    \subfigure[]{\includegraphics[width=0.5\textwidth, angle=90,origin=c]{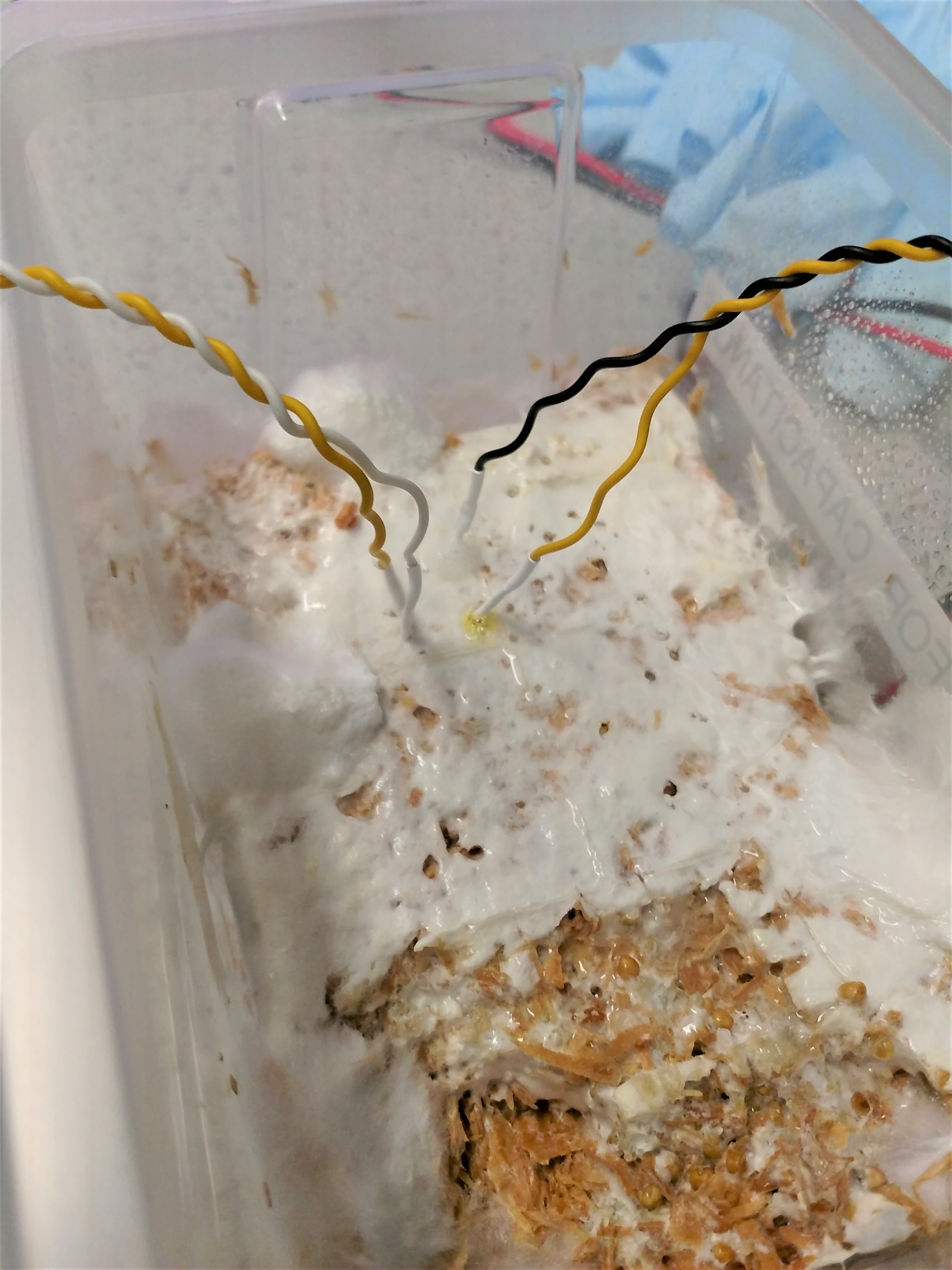} \label{fig:mycelium}
    }
    \subfigure[]{ 
    \includegraphics[width=0.7\textwidth]{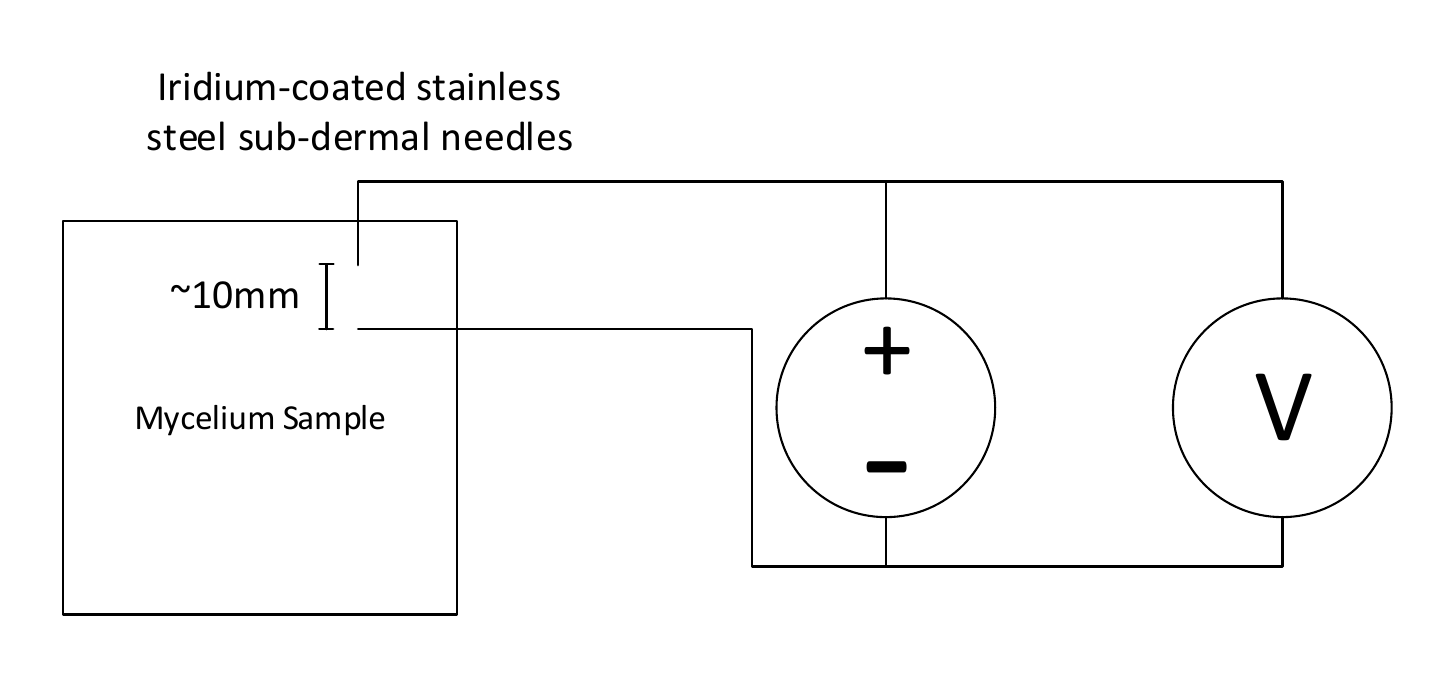}
     \label{fig:discharge_setup}
    }      
    \caption{Experimental setup. (a)~A sample of the mycelium under test. (b)~Voltage discharge measuring set up for mycelium samples}
\end{figure}

Mycelium of the grey oyster fungi \emph{Pleurotus ostreatus} (Ann Miller's Speciality Mushrooms Ltd, UK) was cultivated on damp wood shavings (Fig.~\ref{fig:mycelium}). Control samples of the growth medium, woodshavings, were not colonised by mycelium. Iridium-coated stainless steel sub-dermal needles with twisted cables (Spes Medica SRL, Italy)  were inserted in the colonised substrate. An LCR meter (BK Precision, \emph{model number}) was used to provide a nominal reading of the capacitance of the sample with probes at 10\,mm, 20\,mm, 40\,mm and 50\,mm separation.\par

The  samples were charged using a bench top DC power supply (BK precision 9206) to 50\,V. The power supply output was de-activated and the discharge curve was measured using a bench top digital multi-meter (DMM) Fluke 8846A (Fig.~\ref{fig:discharge_setup}). To fully characterise the capacitance of the samples, both the charge and discharge curves were monitored~\cite{paper:measuring_capacitance} with a number of probe separations (e.g. 10\,mm, 20\,mm, 40\,mm, and 50\,mm). Measurements from the DMM were automated through a serial terminal from a host PC. All bench-top equipment was high impedance to limit power lost through leakage in the test equipment. All plots were generated using MATLAB. The sample interval was approximately 0.33\,s.\par

\section{Results}
\label{sec:results}

\subsection*{Capacitance measurement}

\begin{table}[!tpb]
\caption{Capacitance of mycelium and growth mediums.}
\label{tab:cap}
\begin{center}
\begin{tabular}{c|c|c}
     Sample &  Electrode spacing & Capacitance (pF)\\ \hline
      \multirow{2}{*}{Dry wood shavings} & 10\,mm & 37.6 \\
      & 20\,mm & 37.4 \\ \hline
      \multirow{2}{*}{Damp wood shavings} & 10\,mm & 57 \\ 
      & 20\,mm & 58.6 \\ \hline 
      \multirow{5}{*}{Mycelium (drying)} & 10\,mm & 184 \\
      & 20\,mm & 144 \\
      & 30\,mm & 146 \\ 
      & 40\,mm & 120 \\ 
      & 50\,mm & 118 \\ \hline
        \multirow{5}{*}{Mycelium (freshly watered)} & 10\,mm & 193 \\
      & 20\,mm & 186 \\
      & 30\,mm & 134 \\ 
      & 40\,mm & 125 \\ 
      & 50\,mm & 139 \\ \hline
\end{tabular}
\end{center}
\end{table}

Samples of the growth medium and mycelium were measured for their capacitance using a standard bench top LCR meter (Tab.~\ref{tab:cap}). The measured capacitance value of the mycelium substrate was two to four fold greater than that of the growth medium alone. The values of capacitance were also effected by the moisture content such that, if the capacitance of the mycelium was measured straight after it is sprayed with water, the capacitance typically increased compared to that of dry substrate. \par

\subsection*{Discharge characteristics}

\begin{figure}[!tpb]
    \centering
    \subfigure[]{\includegraphics[width=0.49\textwidth]{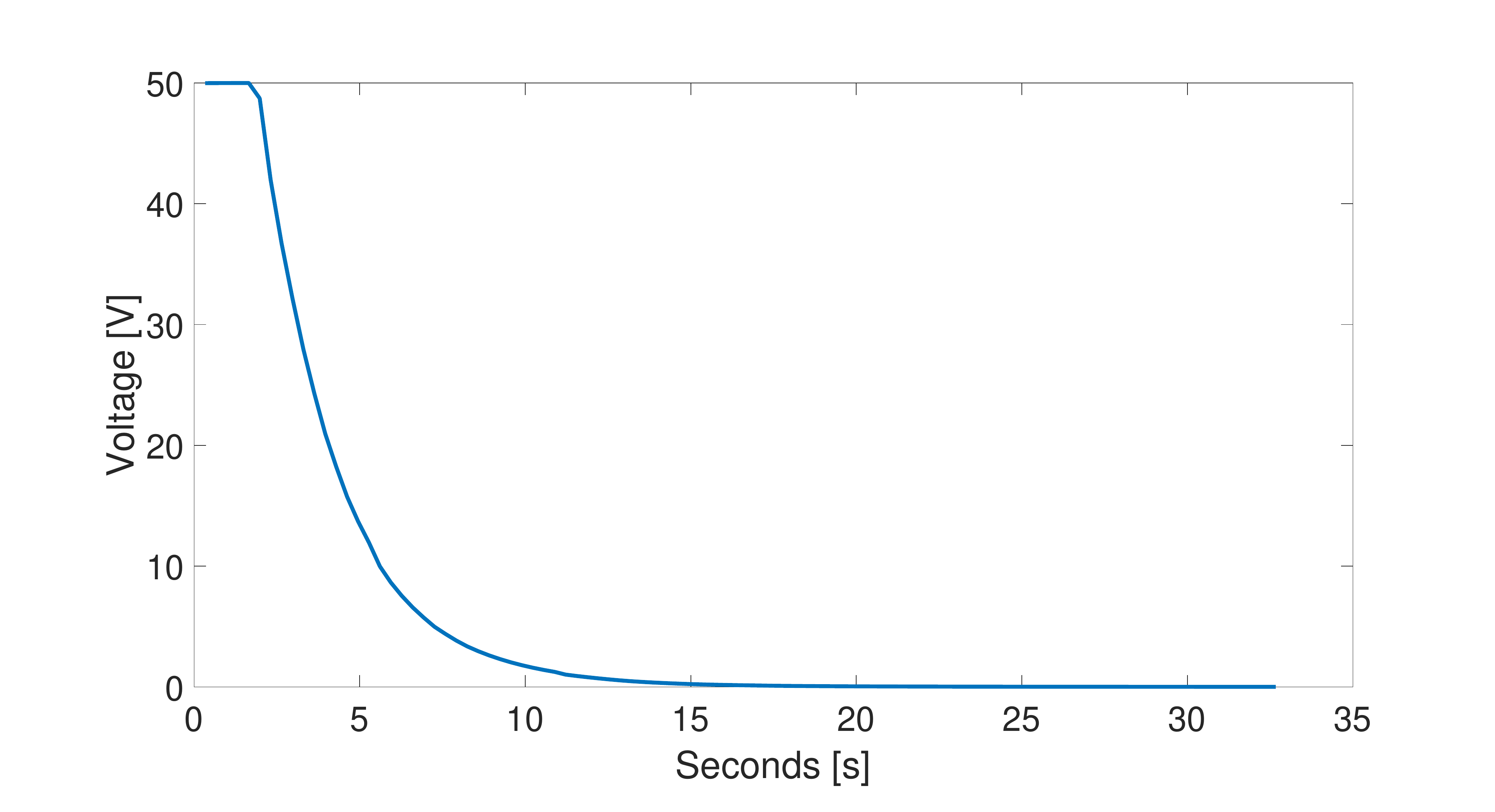}}
    \subfigure[]{\includegraphics[width=0.49\textwidth]{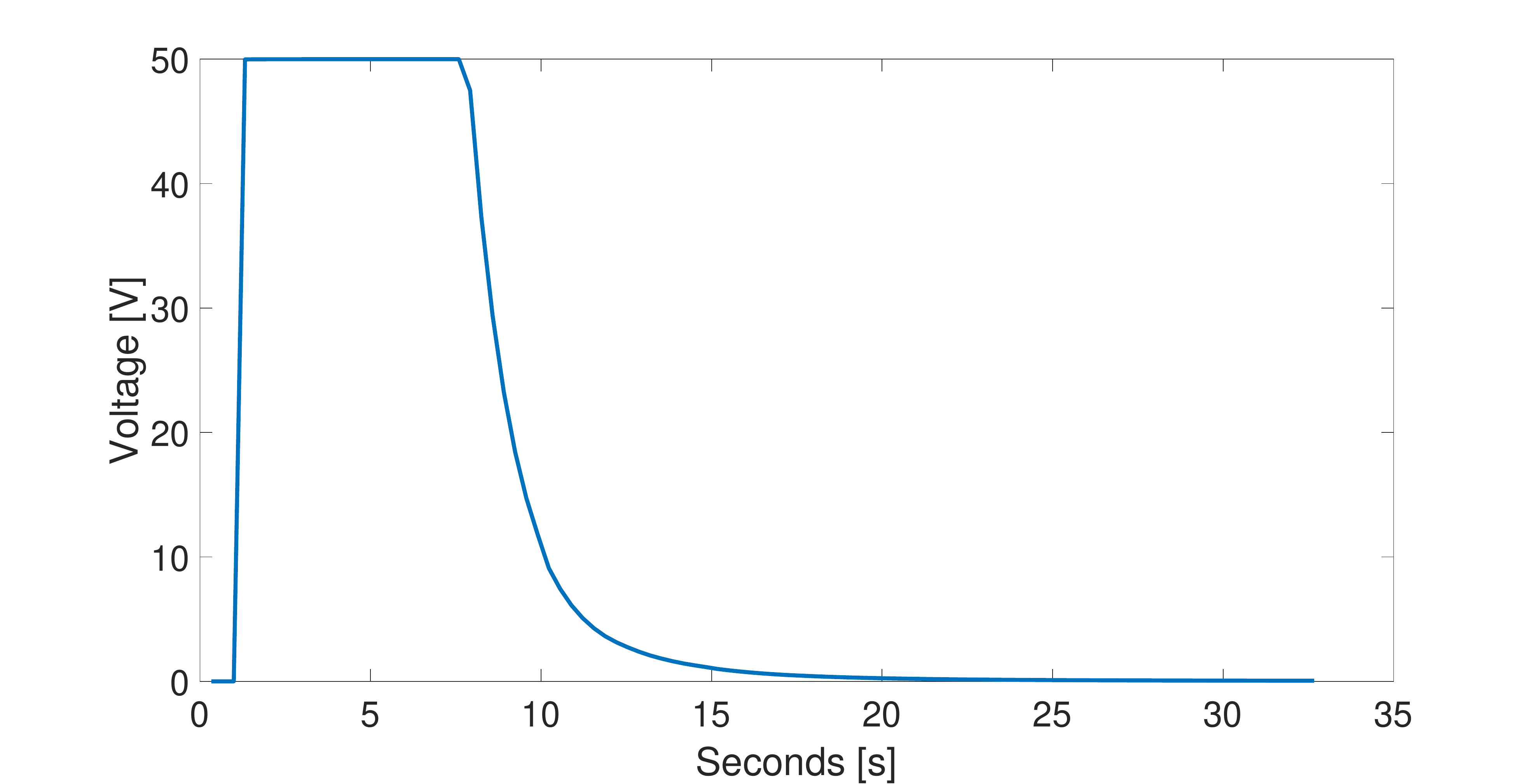}}
    \caption{Discharge of a substrate after being charged to 50\,V with probes separation of 10\,mm. (a) Dry wood shavings. (b) Damp wood shavings --- shavings are immersed in water for half and hour after which excess water is drained. Data are discrete. Line is for eye guidance only.}
    \label{fig:control_discharge}
\end{figure}

\begin{figure}[!tpb]
    \centering
    \includegraphics[width=0.48\textwidth]{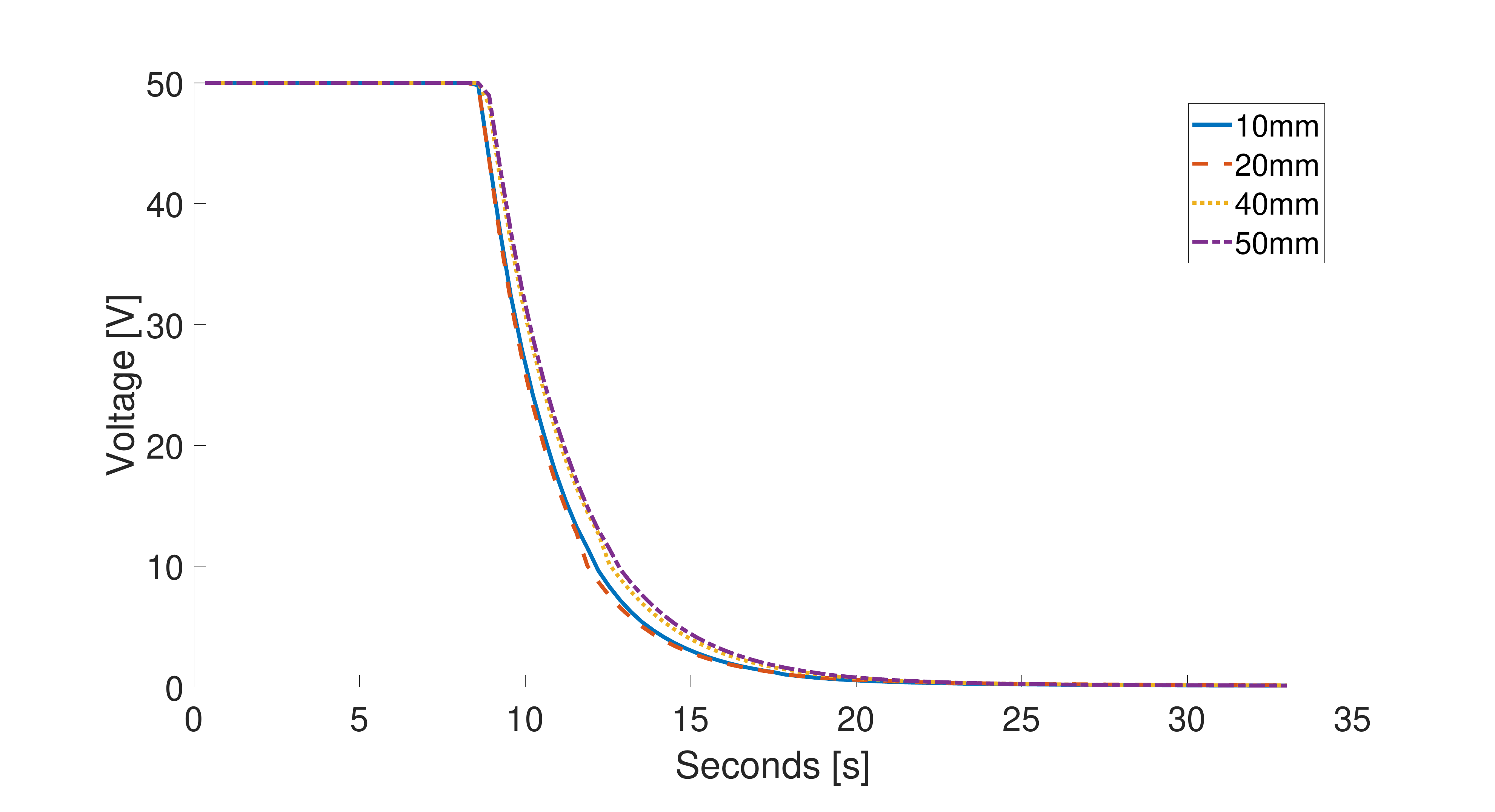}
    \caption{Mycelium sample is charged to 50\,V then allowed to discharge. Electrode spacing is varied (10\,mm, 20\,mm, 40\,mm and 50\,mm). Data are discrete. Line is for eye guidance only.}
    \label{fig:discharge}
\end{figure}

Discharge characteristics of the growth medium and mycelium samples are shown in Figs.~\ref{fig:control_discharge}--\ref{fig:discharge}. Discharge curves were produced by setting up the DC power supply and DMM in parallel with each other. The substrate being tested is then charged to 50\,V and the power supply output is disabled. The DMM continued to periodically measure the remaining charge in the substrate for a period of time. The sample interval was approximately 0.33\,s.\par
The discharge curves for both the growth medium and the mycelium are very steep - approximated by an exponential (\ref{eq:curve_fit}) . 
\begin{equation}
    f(x) = a \cdot e^{b\cdot x}
    \label{eq:curve_fit}
\end{equation}
Where the parameters for 95\% fitness for different mediums can be found in Table~\ref{tab:fitness}.

\begin{table}[!tpb]
\caption{Discharge curve fitness approximation coefficients (with 95\% confidence bounds)}
\label{tab:fitness}
\begin{center}
\begin{tabular}{c|c|c}
    Sample &  $a$ & $b$ \\ \hline
    Dry wood shavings & 104.9  (102.6, 107.2) & -0.4079  (-0.4151, -0.4007)\\ \hline
    Damp wood shavings & 6205  (5164, 7245) & -0.6261  (-0.6464, -0.6058) \\ \hline
    Mycelium w/10\,mm probe separation & 2276  (2198, 2354) & -0.4446  (-0.4481, -0.441) \\ \hline
    Mycelium w/20\,mm probe separation & 2688  (2544, 2833) &  -0.4639  (-0.4695, -0.4583)\\ \hline
    Mycelium w/40\,mm probe separation & 1569  (1458, 1680) & -0.3948  (-0.4021, -0.3875)\\ \hline
    Mycelium w/50\,mm probe separation & 1413  (1311, 1516) &  -0.3817  (-0.3891, -0.3743)\\ \hline
\end{tabular}
\end{center}
\end{table}

The discharge time is governed by equation $\tau = RC$, where $\tau$ is the time constant, $R$ is a resistance, and $C$ is a capacitance.

 With capacitance in the order of pico-Farads, and input impedance of the source and measurement equipment in the order of mega-Ohms, it is expected that the discharge will be in the order of seconds.\par

Comparing the discharge curves of the growth medium to that of the mycelium samples, it is evident that the discharge was not as steep in the mycelium due to the increase in capacitance over the growth medium. However, it was still in the pico-Farad range and, therefore, the majority of charge was lost after just over 5\,s. Increasing the separation distance of the probes (Fig.~\ref{fig:discharge}) had only a minimal effect on the capacitance of the substrate (shown in Tab.~\ref{tab:cap}), and therefore minimal effect on the discharge curve.\par

\begin{figure}[!tpb]
    \centering
    \subfigure[]{\includegraphics[width=0.48\textwidth]{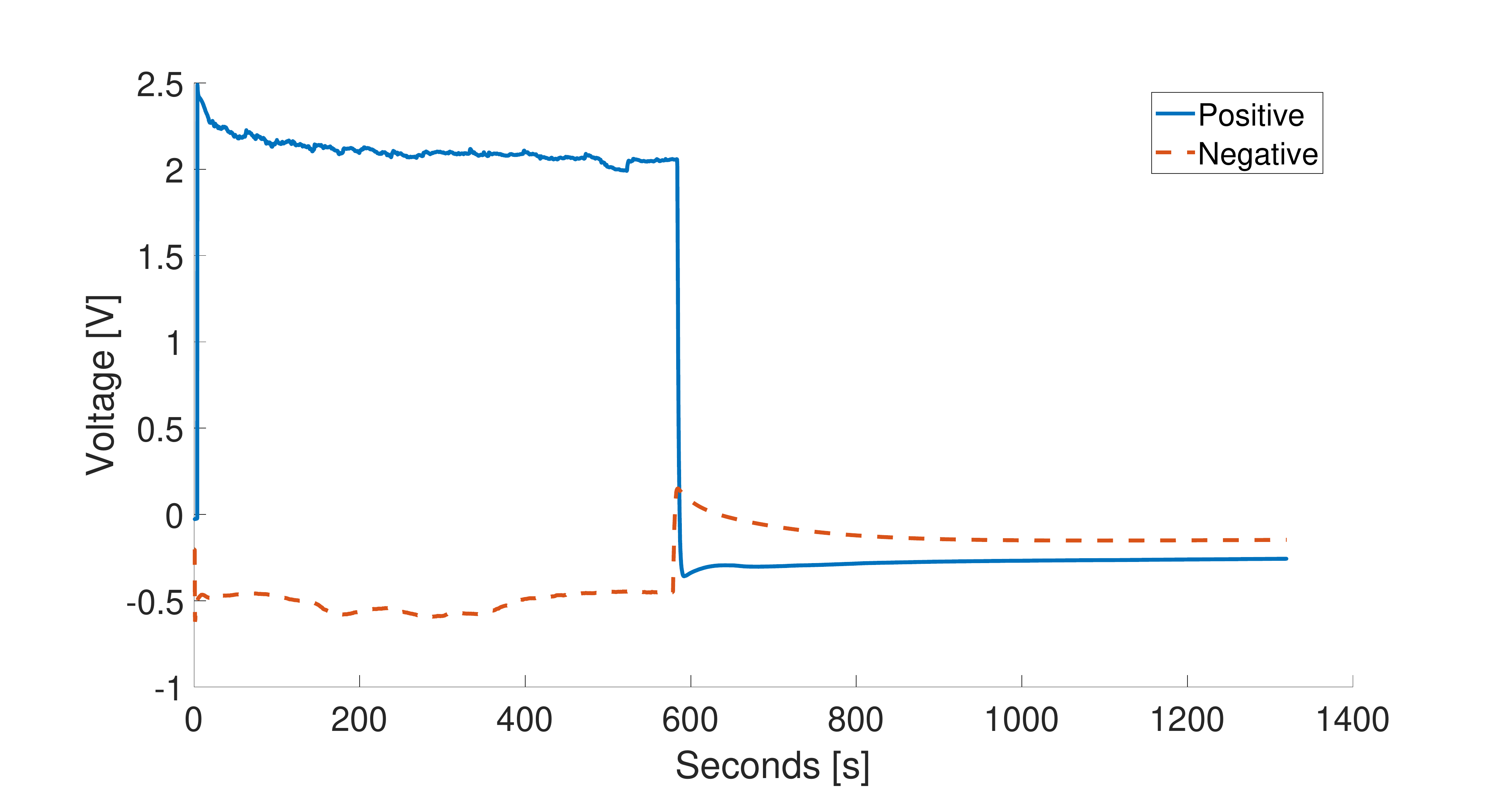}
    \label{fig:series}
    }
    \subfigure[]{\includegraphics[width=0.48\textwidth]{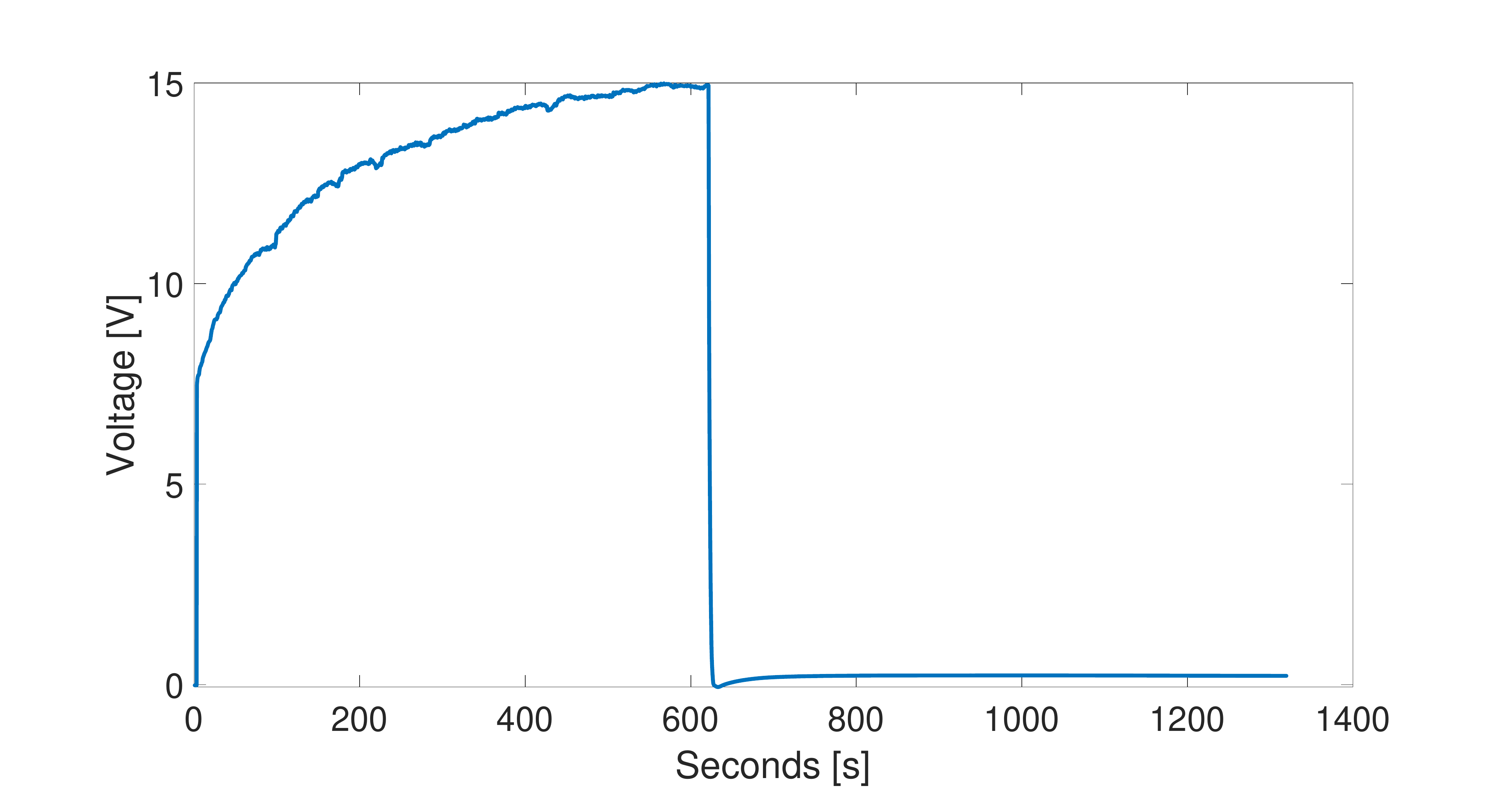}
    \label{fig:parallel}
    }
    \caption{Charging mycelium to 50\,V with source electrodes 10\,mm apart. Substance was charged from approx. 10\,minutes, readings are taken for approx. 22\,minuets in total. (a)~Sense electrodes were placed in series, with the source terminals (10\,mm away from either positive or negative electrodes). (b)~Sense electrodes were in parallel from the source electrodes (10\,mm clearance). Data are discrete. Line is for eye guidance only.}
    \label{fig:discharge_10mm}
\end{figure}

\begin{figure}[!tbp]
    \centering
    \includegraphics[width=0.5\textwidth]{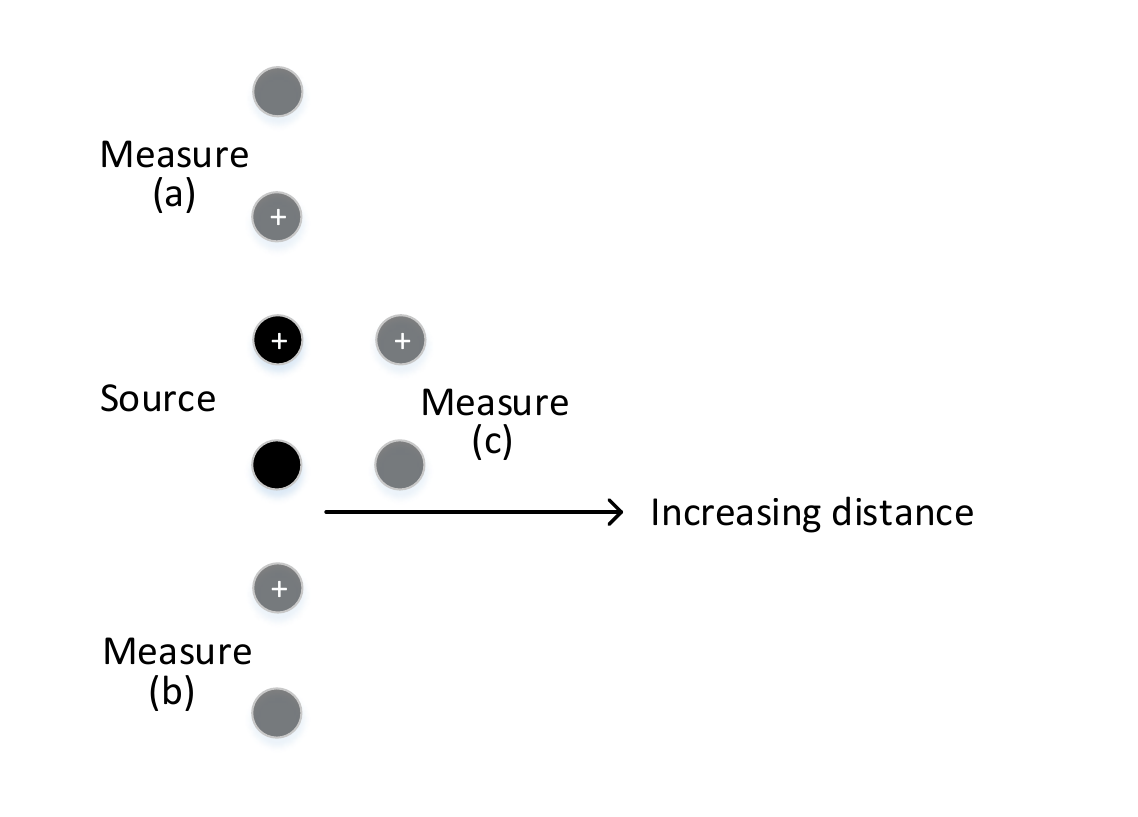}
    \caption{Measurement probes are arranged around the source probes to examine the charge field in the substrate.}
    \label{fig:electrode_arrangement}
\end{figure}

Observing the charging and discharging behaviour of the sample around the source electrodes helps to build a better picture of the current density of the mycelium substrate. Figure~\ref{fig:discharge_10mm} shows how we placed the measurement equipment electrodes 10\,mm away from the source electrodes in the mycelium, in three different locations shown in Fig.~\ref{fig:electrode_arrangement}. The sample was then charged for approximately 10\,mins before the supply was turned off. The electrodes placed in `series' (locations (a) and (b) on Fig.~\ref{fig:electrode_arrangement}) with the charge electrodes (Fig.~\ref{fig:series}) show minimum voltage detected beyond the supply electrodes in the horizontal plane, when the supply is active. Placing the measurement probes in `parallel' (location (c) on Fig.~\ref{fig:electrode_arrangement}) with the supply probes (Fig.~\ref{fig:parallel}) demonstrates the fact that  considerably more current is conducted between the supply probes in the vertical plane. When the supply is de-activated, the voltage around the supplies collapses almost instantly.\par

\subsection*{Charge characteristics}

\begin{figure}[!tpb]
    \centering
    \subfigure[]{\includegraphics[width=0.45\textwidth]{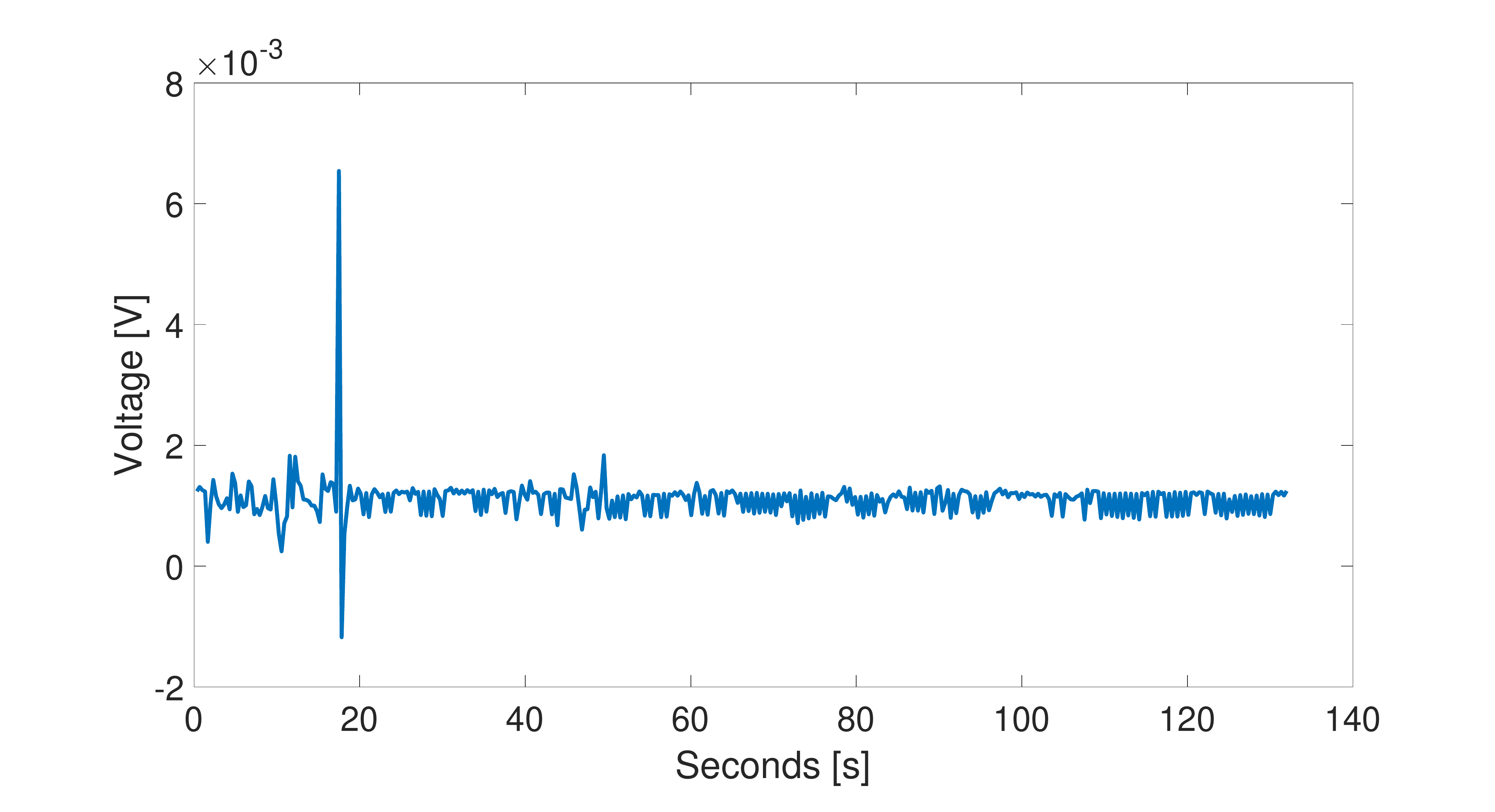}
    \label{fig:dry_parallel}
    }
    \subfigure[]{\includegraphics[width=0.45\textwidth]{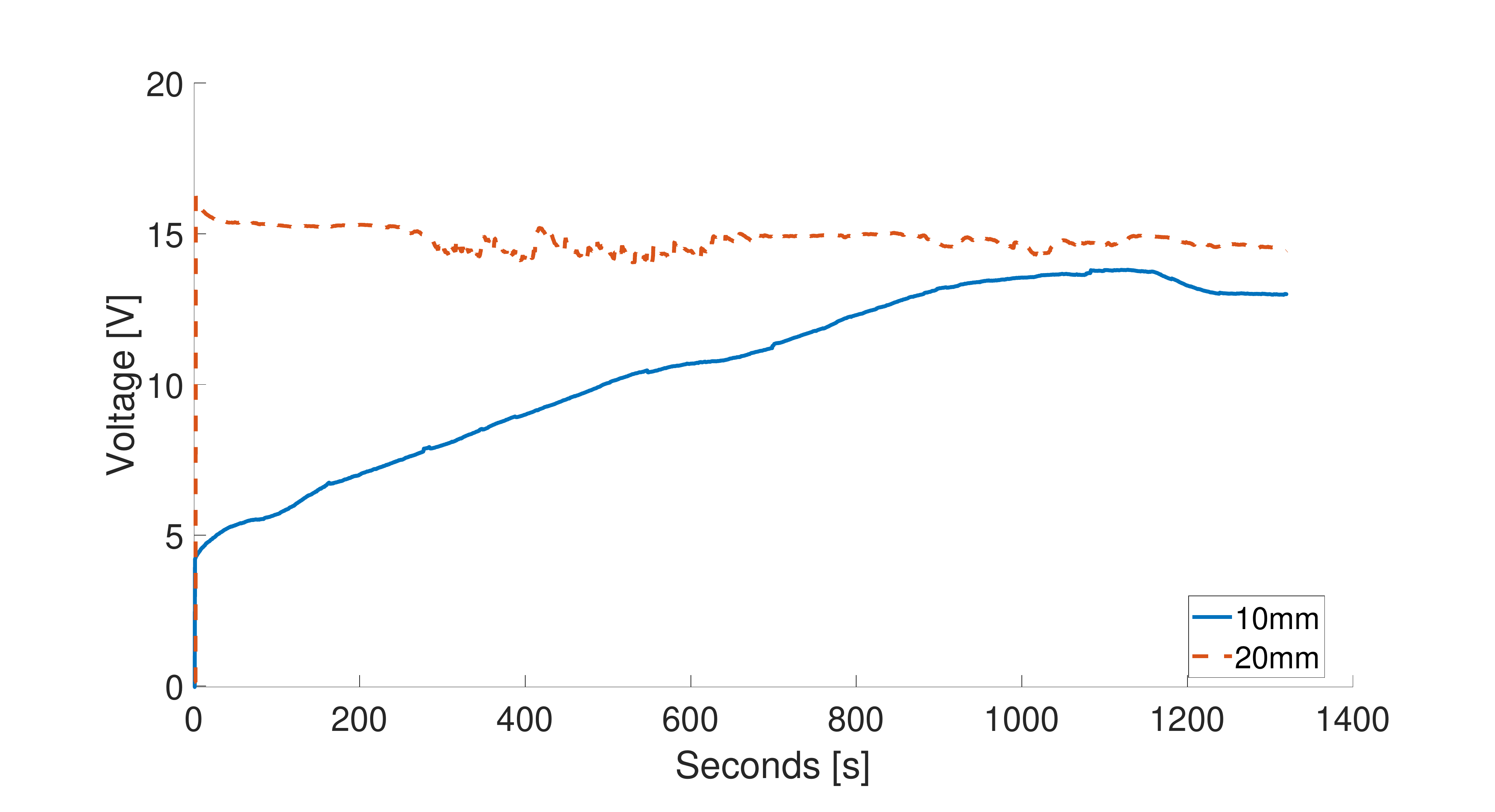}
    \label{fig:damp_parallel}
    }
    \subfigure[]{\includegraphics[width=0.45\textwidth]{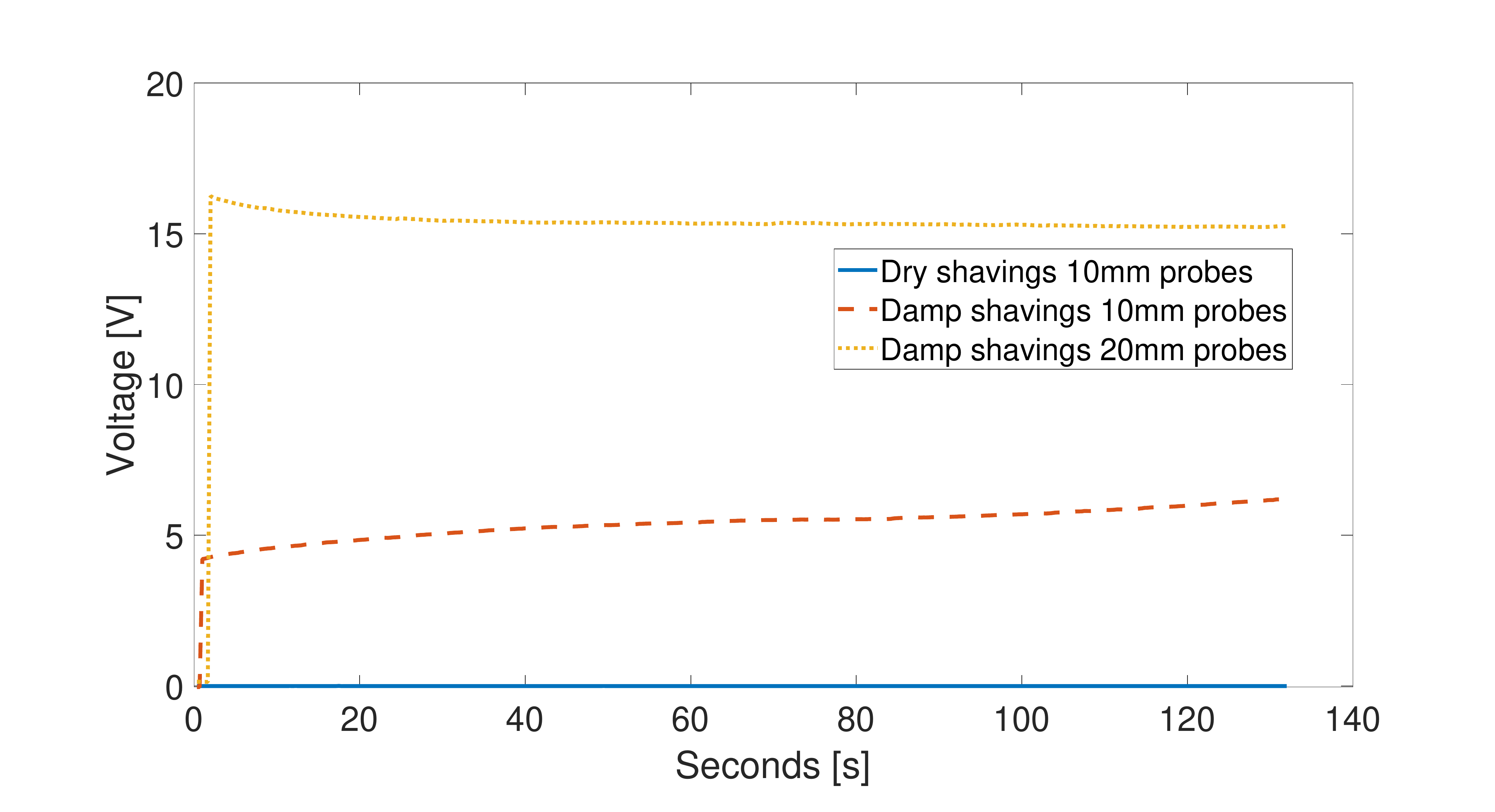}
    \label{fig:damp_and_dry_parallel}
    }
    \caption{Charge dynamics of growth substrate with measurement equipment set in parallel with charge electrodes. Electrode pairs were 10\,mm apart. (a) Dry wood shavings. (b) Damp wood shavings. (c) Dry and damp wood shavings. Data are discrete. Line is for eye guidance only.}
    \label{fig:growth}
\end{figure}

\begin{figure}[!tpb]
    \centering
    \includegraphics[width=0.7\textwidth]{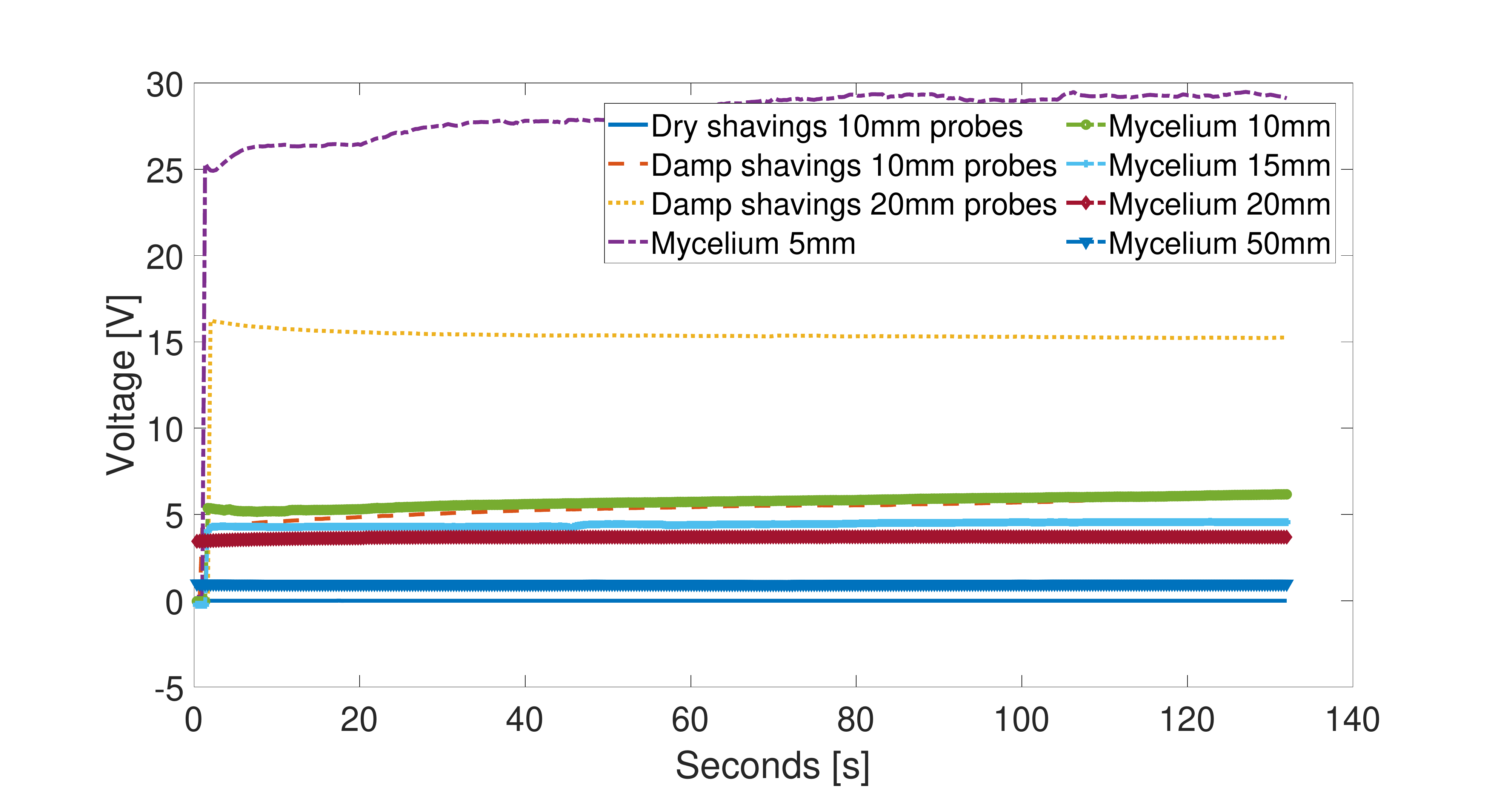}
    \caption{Charge characteristics of wet and dry growth medium and and mycelium with measurement probes at different distances from source probes.}
    \label{fig:all_charge_curves}
\end{figure}

The charge curves in the specimens around the supply electrodes provide an insight into the ability of the substrate to conduct current. The sense electrodes were distanced from the source electrodes by varying amounts. Initially, the growth medium was studied on its own (Fig.~\ref{fig:growth}). The dry wood shavings (Fig.~\ref{fig:dry_parallel}) showed very low voltage across the sense electrodes placed 10\,mm away from the source electrodes (parallel). The dry shavings essentially acted as an open circuit and the measurement electrodes picked up noise. Damp wood shavings form a more contiguous mass and the introduction of the water helped to conduct current (Fig.~\ref{fig:damp_parallel}). From a 50\,V source  a maximum voltage of approx. 15\,V was reached across the measurement electrodes. Although a different charge curve was generated for the two repeated runs with electrodes at different distances, we were primarily interested in the maximum observed voltage over the period Fig.~\ref{fig:damp_and_dry_parallel} superimposes the charge curves from both dry and damp growth medium on to the same axis and Fig.~\ref{fig:all_charge_curves} shows all charge curves for growth medium and mycelium.\par

\begin{figure}[!tpb]
    \centering
    \includegraphics[width=0.7\textwidth]{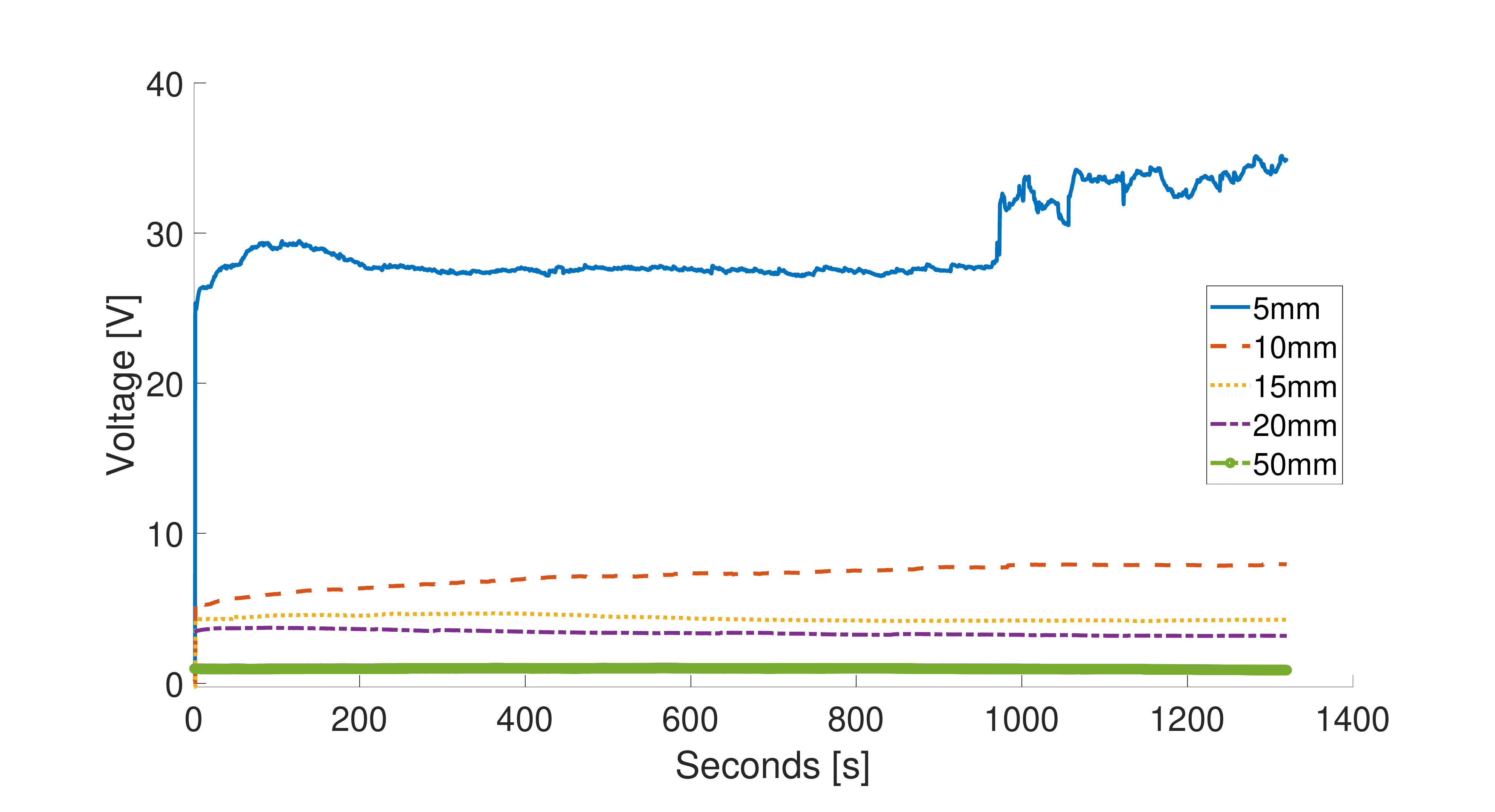}
    \caption{Charging of mycelium sample with measurement equipment set to measure at different distances from supply (5\,mm, 10\,mm, 15\,mm, 20\,mm, and 50\,mm). Supply electrodes and measurement electrodes were arranged in parallel with each other. }
    \label{fig:parallel_measure}
\end{figure}

\begin{figure}[!tpb]
    \centering
    \includegraphics[width=0.7\textwidth]{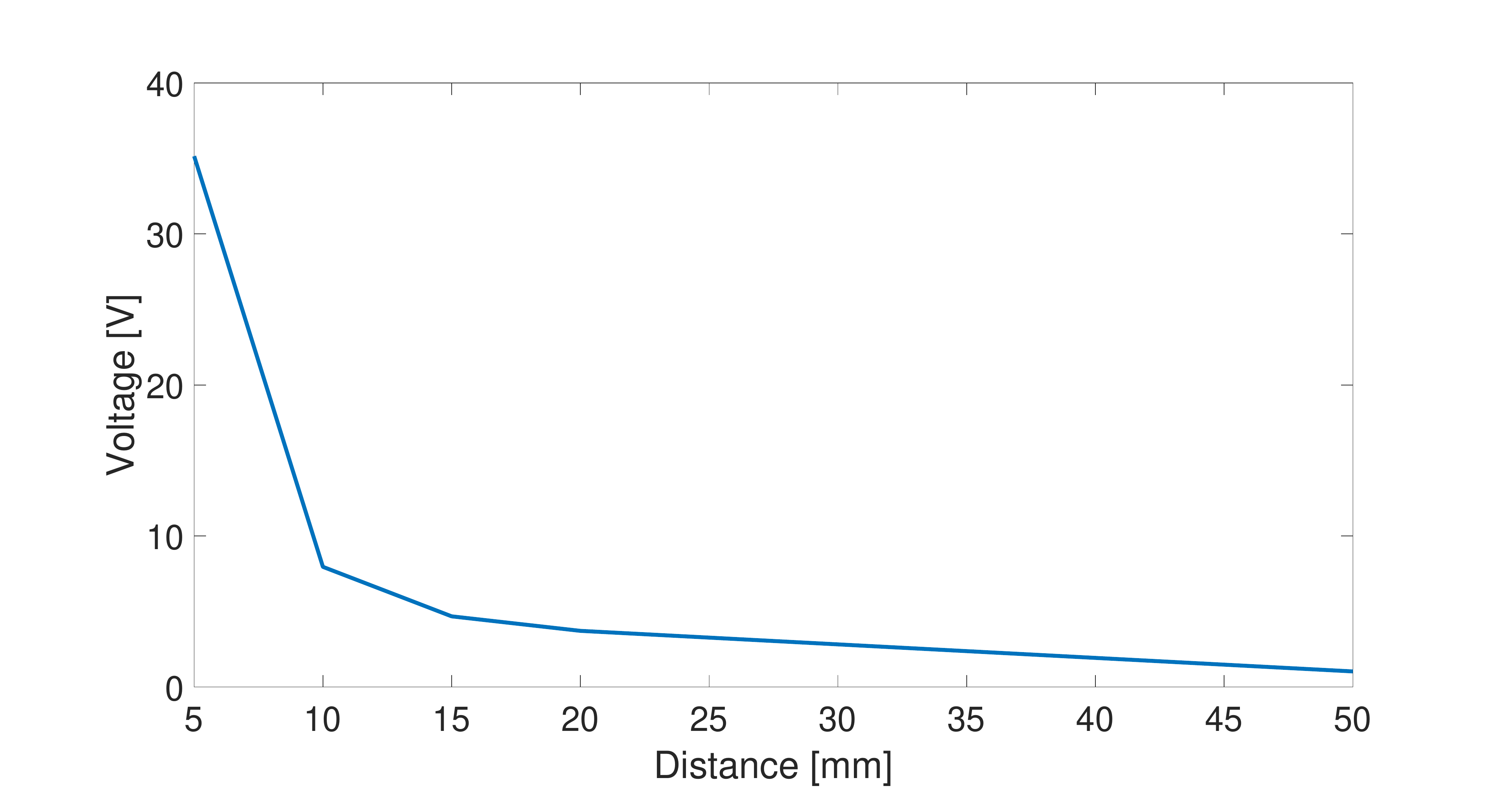}
    \caption{Maximum measured voltage at increasing distance from supply probes in mycelium. Data are discrete. Line is for eye guidance only.}
    \label{fig:max}
\end{figure}

\begin{figure}
    \centering
    \includegraphics[width=0.5\textwidth]{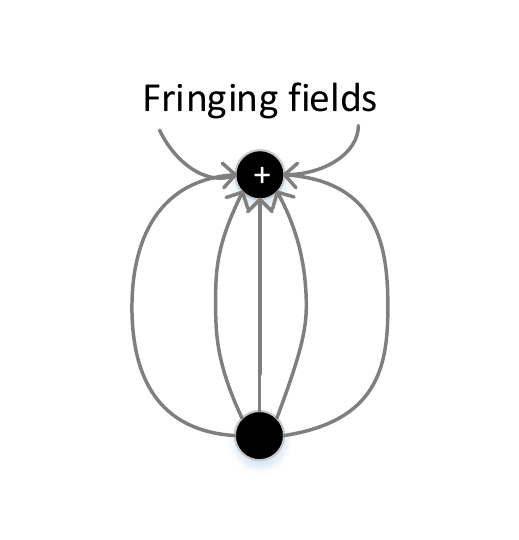}
    \caption{Current density between the two source electrodes. Current flow is shown in the physical direction rather than convention.}
    \label{fig:field}
\end{figure}

Preforming similar experiments with the mycelium substrate (Fig.~\ref{fig:parallel_measure}), it was observed that moving the measurement electrodes further from the supply reduced the measured $V_{\text{max}}$ over the \~22\,min measurement window. Figure~\ref{fig:max} shows that the maximum measured voltage dropped rapidly as the distance from the source electrodes increased. At 10\,mm away from the source, less than $1/5^{th}$ of the supply was measured over 22\,mins, decreasing to less than $1/10^{th}$ at 15\,mm separation. Figure~\ref{fig:field} shows that, for the shortest paths between the two source electrodes, there was a higher current density, indicated by the field lines being closer together. As we move further away from the centre of the two probes, the current density decreased (arrows are shown further apart). Beyond the two probes in the `y--direction', we would expect to experience very little current flow, however there are still fringing field effects which give rise to the small voltages shown in Fig.~\ref{fig:series}.\par

\begin{figure}[!tpb]
    \centering
    \includegraphics[width=0.7\textwidth]{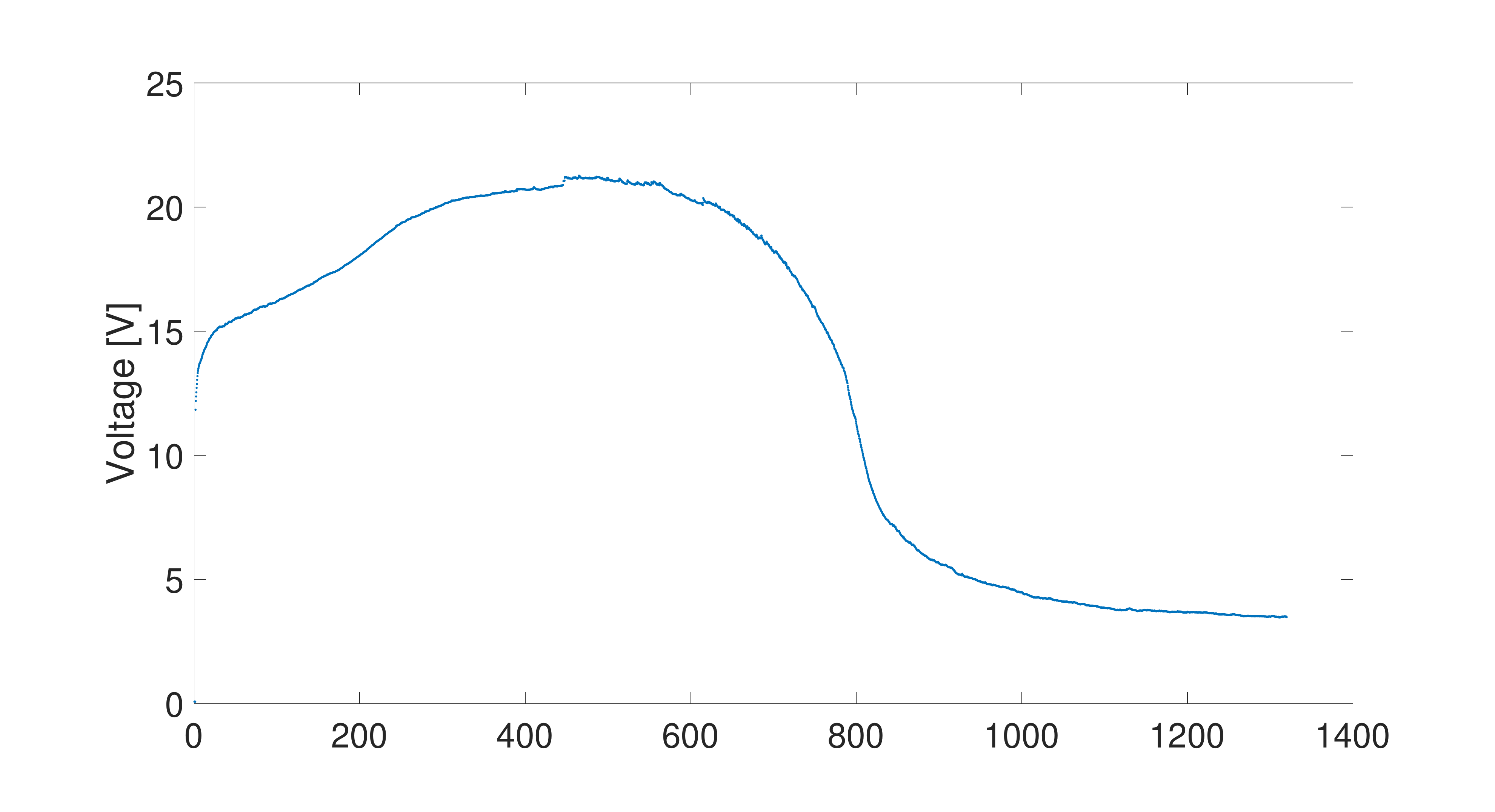}
    \caption{Mycelium charge measured 5\,mm away from 50\,V source while mycelium sample starts to dry out. Data are discrete. Line is for eye guidance only.}
    \label{fig:drying_sample}
\end{figure}

Additionally, the moisture content of the mycelium had an impact on its ability to conduct current. Figure.~\ref{fig:drying_sample} shows that, if the sample of mycelium is continually charged over a period where it is also drying out, the conducted charge can decrease rapidly as the water content vanishes. The measurement electrodes for this sample are  5\,mm away from the source, however we see a decrease in measured voltage of almost 15\,V over the measurement window.\par 

\section{Conclusions}
\label{sec:conclusions}

Mycelium exhibits traditional capacitive characteristics. The capacitance of the substance is in the order of 100's of pico-Farads. Whilst this does not represent vast amounts of storage, it is up to four fold more than that of the growth medium alone. Most noticeably, the charge carrying capability of the substance drops off rapidly when measurements are taken away from the source electrodes. This shows great potential for the use of mycelium networks to conduct or store charge in local `hot spots' that are isolated from other areas in the immediate vicinity. However, it is crucial that the moisture content of the mycelium is kept constant since the ability to carry charge is strongly influenced by moisture content.

Any potential analogue circuits implemented with live mycelium will be vulnerable to environmental conditions, especially humidity, availability of nutrients and removal of metabolites. Ideally, the mycelium networks should be stabilized so they continue functioning whilst drying. This stabilization can be achieved either by coating or priming the mycelium with polyaniline (PANI) or  poly(3,4-ethylenedioxythiophene) and polystyrene sulfonate (PEDOT-PSS). This approach has been proven to be successful in experiments with slime mould \emph{P. polycephaum}~\cite{battistoni2017organic,battistoni2017organic,cifarelli2015bio}, and thus it is likely that a similar technique may be applied to fungi. Moreover, PABI and PEDOT-PSS incorporated in, or interfaced with, mycelium can bring additional functionality in terms of conductive pathways~\cite{yoo2008narrowing}, memory switches~\cite{howard2013spice,demin2015electrochemical} and synaptic-like learning~\cite{berzina2009optimization,lapkin2018spike}. An optional route toward the functional fixation of mycelium would be doping the networks with substances that affect the electrical properties of mycelium, such as carbon nanotubes, graphene oxide, aluminium oxide, calcium phosphate. Similar studies conducted in our laboratory using slime mould and plants have shown that such an approach is feasible~\cite{gizzie2016hybridising,Gizzie2016}. Moreover using a combination of PANI and carbon nanotubes in the mycelium network afford it supercapacitive properties~\cite{dong2007preparation,frackowiak2006supercapacitors}. Another potential direction of future studies would be to increase the capacity of the mycelium as a result of modifying the network geometry by varying nutritional conditions and temperature~\cite{boddy1999fractal, hoa2015effects, rayner1991challenge,regalado1996origins}, concentration of nutrients~\cite{ritz1995growth} or with chemical and physical stimuli~\cite{bahn2007sensing}. With regards to the impact of our finding for the field of unconventional computing, we believe further research on experimental laboratory implementation of capacitive threshold logic~\cite{ozdemir1996capacitive,medina2018reconfigurable}, adiabatic capacitive logic~\cite{pillonnet2017adiabatic} and capacitive neuromorphic architectures~\cite{wang2018capacitive} will yield fruitful insights.

\section*{Acknowledgements}
This project has received funding from the European Union's Horizon 2020 research and innovation programme FET OPEN ``Challenging current thinking'' under grant agreement No 858132.

\section*{Author contributions}
A.B. conceived the idea of experiments. A.A. and A.P. prepared the substrate colonised by mycelium. A.B. performed experiments, collected data and produced all plots in the manuscript.
A.B. and A.A. prepared manuscript (wrote and reviewed all contents).
A.P. reviewed manuscript.
\bibliographystyle{plain}
\bibliography{references_capacitors, references, references_photosensor,plant,fungalbib}

\end{document}